\documentclass[runningheads]{llncs}
\usepackage{graphicx}
\usepackage{amsmath,amssymb}
\usepackage{xspace}
\usepackage{multirow,setspace,verbatim,amsfonts,amsmath,amsbsy,epsfig,url}
\usepackage{gensymb}
\usepackage{booktabs}
\usepackage{cite}
\usepackage{makecell} 
\usepackage{epstopdf}
\usepackage{array,multirow}
\newcolumntype{P}[1]{>{\centering\arraybackslash}p{#1}}
\newcolumntype{M}[1]{>{\centering\arraybackslash}m{#1}}

\usepackage{algorithm}
\usepackage[noend]{algpseudocode}
\usepackage{tikz}
\usepackage{mathrsfs} 
\usepackage[algo2e]{algorithm2e} 
\usepackage{array}
\usepackage{color}
\usepackage{tabularx}

\begin{document}
%
\title{Vision Transformer for COVID-19 CXR Diagnosis using  Chest X-ray Feature Corpus}
\titlerunning{ViT for COVID-19 Diagnosis using Low-level CXR Feature Corpus}
%
\author{Sangjoon Park\inst{1}\orcidID{0000-0002-9223-3172}\and
Gwanghyun Kim\inst{1}\and
Yujin Oh\inst{1}\and
Joon Beom Seo\inst{2}\and
Sang Min Lee\inst{2}\and
Jin Hwan Kim\inst{3}\and
Sungjun Moon\inst{4}\and
Jae-Kwang Lim\inst{5}\and
Jong Chul Ye\inst{1}\orcidID{0000-0001-9763-9609}}
\authorrunning{S.Park et al.}
%
\institute{Korea Advanced Institute of Science and Technology, Daejeon, South Korea \\
\email{\{depecher, gwang.kim, yujin.oh, jong.ye\}@kaist.ac.kr}\and
{Asan Medical Center, University of Ulsan College of Medicine, Seoul, South Korea} \\ \and
{College of Medicine, Chungnam National Univerity, Daejeon, South Korea} \\ \and
{College of Medicine, Yeungnam University, Daegu, South Korea} \\ \and
{School of Medicine, Kyungpook National University, Daegu, South Korea}}
\maketitle              
\begin{abstract}
Under the global COVID-19 crisis, developing robust diagnosis algorithm for COVID-19 using CXR is hampered by the lack of the well-curated COVID-19 data set, although CXR data with other disease are abundant. This situation is suitable for vision transformer architecture that can exploit the abundant unlabeled data using  pre-training. However, the direct use of existing vision transformer that uses the corpus generated by the ResNet is not optimal for correct feature embedding. To mitigate this problem, we propose a novel vision Transformer by using the low-level CXR feature corpus that are obtained  to extract the abnormal CXR features. Specifically, the backbone network is trained using large public datasets to obtain the abnormal features in routine diagnosis such as consolidation, glass-grass opacity (GGO), etc. 
Then, the embedded features from the backbone network are used as corpus for vision transformer training.
We examine our model on various external test datasets acquired from totally different institutions to assess the generalization ability. Our experiments demonstrate that our method achieved the state-of-art performance and has better generalization capability, which are crucial for a widespread deployment.

\end{abstract}
\section{Introduction}


The novel coronavirus disease 2019 (COVID-19), caused by severe acute respiratory syndrome coronavirus-2, is an ongoing pandemic resulting in 113,695,296 people infected with 2,526,007 death worldwide as of 1 March 2021.
In the face of the unprecedent pandemic by COVID-19, public health care systems have confronted challenges in many aspects including critical shortage of medical resources, while many health care provider have themselves been infected \cite{ng2020covid}. Because of its highly transmissible and pathologic natures of COVID-19, the early screening of COVID-19 is becoming increasingly important to prevent further spread of disease and lessen the burden of health care systems.

Currently, real-time polymerase chain reaction (RT-PCR) is the gold standard in COVID-19 confirmation due to high sensitivity and specificity \cite{tahamtan2020real}, but it takes several hours to get the result. As many patients with confirmed COVID-19 present radiological findings of pneumonia, radiologic examinations may be useful for fast diagnosis \cite{shi2020radiological}. Though chest computed tomography (CT) has superior sensitivity and specificity for diagnosis of COVID-19 \cite{bernheim2020chest}, the routine use of CT places a huge burden on health care system due to its high cost and relatively longer scan time than chest radiograph (CXR). Therefore, there exist practical advantages to use CXR as primary screening tool under global pandemic. The common CXR findings of COVID-19 include bilateral involvement, peripheral and lower zone dominance of ground glass opacities and patchy consolidations \cite{cozzi2020chest}. Even though it has been reported that the sensitivity and specificity of COVID-19 diagnosis with CXR alone is lower than with CT or RT-PCR \cite{wong2020frequency}, CXR still has its potential for the fast screening of COVID-19 during the patient triage, determining the priority of patient's care to help saturated health care system in pandemic situation.

Accordingly, many approaches have been proposed using deep learning to diagnose COVID-19 with CXR \cite{wang2020covid, hemdan2020covidx, narin2020automatic, oh2020deep}, but they suffered from common problems of limited number of labelled COVID-19 data, resulting poor generalization ability \cite{hu2020challenges, zech2018variable}. The reliable generalization performance on unseen, totally different data set is crucial for real world adoption of the system.

 In general, the most common approach to solve this problem is to build adversarially robust model with millions of training data \cite{chen2020more}. However, constructing well-curated dataset containing large number of labelled COVID-19 cases is difficult due to saturation of health care system in many countries. Though the previous studies have tried to mitigate the problem either by using transfer learning from other large dataset like ImageNet \cite{apostolopoulos2020covid}, or by utilizing weakly-supervised learning method \cite{zheng2020deep, wang2020weakly} and anomaly detection \cite{zhang2020covid}, their performances are often suboptimal and do not guarantee the ability to generalize. In addition, as COVID-19 usually involves both lung fields with lower zone dominance, the model should extract features based on the global manifestation of the diseases.
 

Transformer, which was first introduced in the field of natural language processing (NLP), is a deep neural network based on self-attention mechanism that results in significantly large receptive fields \cite{vaswani2017attention}. After achieving astounding results in NLP, it has inspired the vision community to study its applications in computer vision since it enables modeling long-range dependency within images. Vision Transformer (ViT) has first showed how Transformer can totally replace standard convolution operations in deep neural network achieving state-of-the-art (SOTA) performance \cite{dosovitskiy2020image}. However, training Vision Transformer from scratch requires huge amount of data, so that  the authors also suggested hybrid model by conjugating conventional convolutional neural network (e.g. ResNet) backbone that produces initial feature embedding. As such, Transformer,  being trained using the feature corpus generated by the ResNet backbone,
can mainly focus  on learning the global attention.
Empirical results shows that   the hybrid model present better performance in small-sized data set.

Although this preliminary results is promising,
there are still remaining concerns that the corpus generated by the ResNet may not be an optimal input feature embedding for diagnosis with CXRs.
Fortunately, there are several publicly available large-scale datasets for CXR classification which was built before the COVID-19 outbreak.
 Among them CheXpert \cite{irvin2019chexpert} dataset consist of labeled abnormal observations including low-level CXR features (e.g. opacity, consolidation, edema, etc.) useful for diagnosis of infectious diseases.  Furthermore, there exists many advanced CNN architectures  including 
the model proposed by \cite{ye2020weakly}, which utilizes probabilistic class activation map (PCAM) pooling to explicitly leverage the benefit of class activation map to improve both classification and localization ability for these low-level features. 
Therefore, we propose a novel vision Transformer which  utilizes this existing network  as a backbone for CXR low-level feature embedding, upon which vision Transformer
is trained using the generated corpus from the backbone network.

It is remarkable that  our network fundamentally resembles the text classification task using Transformer, which classifies the entire sentence by aggregating the meaning, location and relationship of words in a sentence, using the low-level word embedding.
Furthermore,  our network emulates the clinical experts who make a CXR image-level diagnosis (e.g. normal, tuberculosis, COVID-19 pneumonia, etc.) by collating the information of low-level features in terms of their pattern, multiplicity, distribution and locations.

\subsubsection{Contribution.} In this paper, we proposed a Vision Transformer model tailored for COVID-19 CXR diagnosis, leveraging the low-level CXR feature corpus attained from prebuilt large-scale public dataset for CXR features such as opacity, consolidation, edema, etc. 
We show that our method is superiority to both baseline Vision Transformer and SOTA models especially in terms of the generalization performance on unseen datasets. We also provided clinically interpretable visualization results of model, which is of great help for COVID-19 diagnosis and localization.

\section{Method}

\begin{figure}[!t]
\centering
\includegraphics[width=\textwidth]{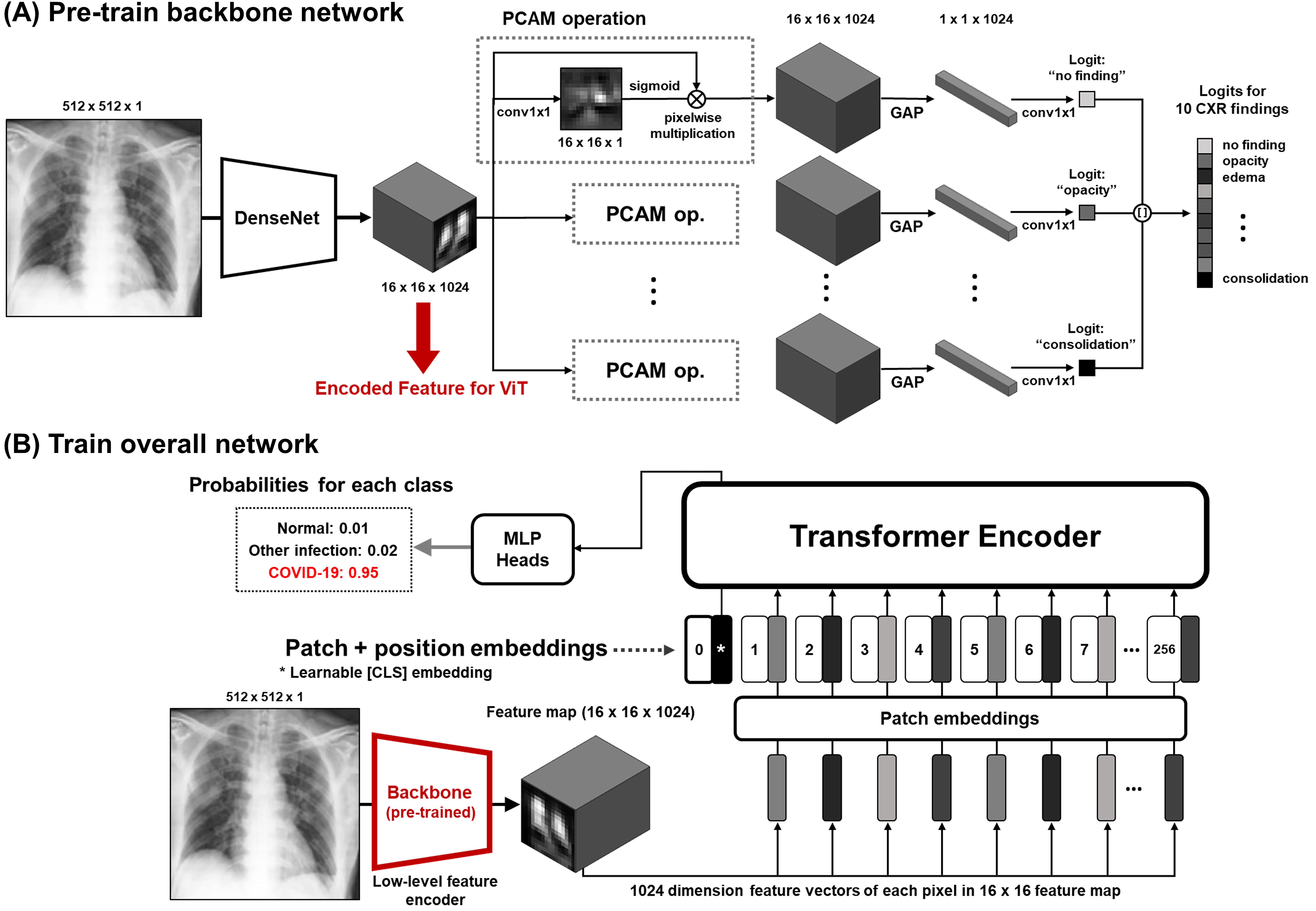}
\caption{Overall framework of our method. (A) We first trained the backbone network utilizing probabilistic-CAM (PCAM) pooling for each of ten low-level CXR findings, and then trained (B) the overall network, which consists of backbone and Transformer.} \label{fig1}
\end{figure}

The merit of our model is that Transformer can exploit the low-level CXR feature corpus, which was obtained from the backbone network trained to extract abnormal CXR features from publicly available large and well-curated CXR dataset.
The overall framework of our method is illustrated in Fig. \ref{fig1}.


\subsubsection{Pre-training Backbone Network for Low-level Feature Corpus.} As a backbone network to extract low-level CXR feature corpus from an image, we adopted a modified version of model proposed by \cite{ye2020weakly}, which utilizes probabilistic class activation map (PCAM) pooling to explicitly leverage the benefit of class activation map to improve both classification and localization ability (see Fig. \ref{fig1} (A)). The backbone network was trained beforehand with prebuilt public CXR dataset to classify 
10 labelled observations including no finding, cardiomegaly, lung opacity, edema, consolidation, pneumonia, atelectasis, pneumothorax, pleural effusion and support devices.

As shown in Fig.~\ref{fig1} (A), there are several layers one could extract the feature embedding, and we found that the intermediate level
embedding before the PCAM operation contains the most useful information. However, care should be taken since the PCAM unit trained with specific low-level CXR features (e.g.cardiomegaly, lung opacity, edema, consolidation) was essential to improve the accuracy of the intermediate level feature embedding by guiding the feature aligned to provide the optimal PCAM maps.
Specifically, with the pre-trained backbone network $\boldsymbol{G}$, an input image $\boldsymbol{x} \in \mathbb{R}^{{H}\times{W}\times{C}}$ is encoded into intermediate feature map $\boldsymbol{f} \in \mathbb{R}^{{H'}\times{W'}\times{C'}}$. We used the $C'$ dimension feature vectors $f$ of each ${H'}\times{W'}$ pixels as encoded representations for low-level features at each pixel locations, and constructed the low-level CXR feature corpus.

\begin{flalign}
&&\boldsymbol{f} &= \boldsymbol{G}(\boldsymbol{x}), & \boldsymbol{x} \in \mathbb{R}^{{H}\times{W}\times{C}}, \; \boldsymbol{f} \in \mathbb{R}^{{H'}\times{W'}\times{C'}}\\
&&\boldsymbol{f} &= [f^{1};f^{2};f^{3};...;f^{{H'}\times{W'}}], & f \in \mathbb{R}^{{C'}}
\end{flalign}

\subsubsection{Vision Transformer.} Similar to BERT \cite{devlin2018bert}, Vision Transformer model is encoder-only architecture (see Fig. \ref{fig1} (B)). As the Transformer encoder uses constant latent vector of dimension $D$, we first projected encoded features $f$ of dimension $C'$ into $f_{p}$ of dimension $D$ using $1$ $\times$ $1$ convolution kernel. Similar to \texttt{[class]} token of BERT, we prepended additional learnable embedding vector $f_\texttt{cls}\;$ to projected features $\boldsymbol{f}_{p}$, to make the last $L$ layer output of this \texttt{[class]} token $z^{0}_{L}$ represent the diagnosis of a whole CXR image (= $y$) by attaching the classification heads to $z^{0}_{L}$. In addition, we added a positional embedding $\textbf{E}_{pos}$ to encode a notion of the sequential order  to the projected features $\boldsymbol{f}_{p}$. The Transformer encoder layers used in our model are the same as standard Transformer encoder consisting of alternating layers of multihead self-attention (MSA), multilayer perceptron (MLP), layer normalization (LN), and residual connections in each block, which is described as follows:

\begin{flalign}
&&[f_{p}^{1};f_{p}^{2};f_{p}^{3};...;f_{p}^{{H'}\times{W'}}] &= \texttt{conv}([f^{1};f^{2};f^{3};...;f^{{H'}\times{W'}}]), &f_{p} \in \mathbb{R}^{{D}}\\
&&[z^{0}_0;z^{1}_0;z^{2}_0;...;z^{{H'}\times{W'}}_0] &= [f_\texttt{cls};f_{p}^{1};f_{p}^{2};f_{p}^{3};...;f_{p}^{{H'}\times{W'}}] + \textbf{E}_{pos}\\
&&\textbf{z}_0 &= [z^{0}_0;z^{1}_0;z^{2}_0;...;z^{{H'}\times{W'}}_0]\\
&&\textbf{z}'_{l} &= \texttt{MSA}(\texttt{LN}(\textbf{z}_{l-1})) + \textbf{z}_{l-1}, &l = 1 ... L\\
&&\textbf{z}_{l} &= \texttt{MLP}(\texttt{LN}(\textbf{z}'_{l})) + \textbf{z}'_{l}, &l = 1 ... L\\
&&y &= \texttt{HEAD}(z_{L}^{0})
\end{flalign}

\subsubsection{Model Interpretability.} For model interpretability, we adopted a saliency map visualization method tailored for Vision Transformer. Different from other methods relying on attention maps or heuristic propagation of attention, the method proposed by \cite{chefer2020transformer} assigns local relevance with deep Taylor decomposition, and propagate the local relevance throughout the layers. By the relevance propagation, this method overcomes the challenges of attentions layers and the skip connections. 

\section{Experimental Results}


\subsubsection{Datasets and Partitioning.} For the pre-training of the feature encoder network, we used CheXpert \cite{irvin2019chexpert} dataset containing 10 labelled observations including no finding, cardiomegaly, lung opacity, edema, consolidation, pneumonia, atelectasis, pneumothorax, pleural effusion and support devices. After excluding lateral view images, 190,847 CXRs were available. For training and evaluation of Transformer model for COVID-19 diagnosis using CXR, we aggregated retrospectively collected CXR data from four independent local institutions (Asan Medical Center [AMC], Seoul, Korea; Chonnam National Univerity Hospital [CNUH], Daejeon, Korea; Yeungnam University Hospital [YNU], Daegu, Korea; Kyungpook National University Hospital [KNUH], Daegu, Korea), which were labelled by board-certified radiologists and publicly available CXR datasets containing labels of infectious disease (Brixia \cite{signoroni2020end}, BIMCV \cite{de2020bimcv}, NIH \cite{wang2017chestx} datasets). Total dataset was curated into three classes; normal, other infections (which includes bacterial pneumonia and tuberculosis), and COVID-19. The numbers of images for each class are summarized in Table \ref{table:dataset}. To evaluate the generalization ability of models in various institutional settings, we set aside 3 institutional datasets (CNUH, KNUH, YNU) collected from totally different institutions with different devices and settings as external test datasets as they contain the cases of all three label classes.

\begin{table}[!t]
  \caption{Summary of dataset resources and disease classes.}
  \label{table:dataset}
  \centering
  \resizebox{\textwidth}{!}
  {%
  \begin{tabular}{M{2cm}|M{1.5cm}|M{1.5cm}|M{1.5cm}|M{1.5cm}|M{1.5cm}|M{1.5cm}|M{1.5cm}|M{1.5cm}}
  \hline
  \multirow{2}{*} {\textbf{Resource}} & \multirow{2}{*} {\textbf{Total}} & \multicolumn{3}{c|}{\textbf{External test data}} & \multicolumn{4}{c}{\textbf{Training and Validation data}}\\ 
  \cline{3-9}
  & & {CNUH} & {YNU}& {KNUH}  & {AMC} & {NIH} & {Brixia} & {BIMCV} \\
  \cline{2-3} \cline{6-7}
  \hline
   \multicolumn{1}{l|}{Normal} & 13,649 & 320 & 300 & 400 & 8,861 & 3,768 & {-} & {-}\\
  \multicolumn{1}{l|}{Other infection} & 1,468 & 39 & 144 & 308 & 977 & {-} & {-} & {-} \\
   \multicolumn{1}{l|}{{COVID-19}}  & 2,431 & 6 & 8 & 80 & {-} & {-} & 1,929 & 408\\
   \cline{1-9}
   \multicolumn{1}{l|}{{Total images}}  & 17,548 & 365 & 452 & 788 & 9,838 & 3,768 & 1,929 & 408\\
   \hline
  \end{tabular}}
\end{table}

\subsubsection{Implementation Details and Evaluation Metrics.} Our pre-processing method includes histogram equalization, Gaussian blurring with $3 \times 3$ kernel, normalization and resizing to $512 \times 512$.
As the backbone network, we took up on the network architecture that scored the first place in CheXpert challenge 2019 \cite{ye2020weakly}, which consists of DenseNet-121 backbone followed by PCAM operations for each class. In particular, we used intermediate feature map of size $16 \times 16 \times 1024$ before PCAM operation as input for Transformer, since this feature map contains a common representation of all the abnormal findings. As this feature map already encodes the representations for important CXR findings, we adopted relatively simple transformer architecture with four encoder layers with eight attention heads. For training of backbone network, we used Adam optimizer with learning rate of 0.0001. The backbone network was trained for 160,000 optimization steps with step decay scheduler. The batch size was set to 8.
For training of model for COVID-19 diagnosis, We used SGD optimizer (momentum = 0.9) with max grad norm set to 1 with learning rate of 0.001. The model was trained for 10,000 optimization steps with cosine warm-up scheduler (warm-up steps = 500), with batch size set to 8. These optimal hyperparameters are determined experimentally. We adopted area under the ROC curve (AUC) as our evaluation metric, but also calculated sensitivity, specificity and accuracy after adjusting the threshold to meet the sensitivity $\geq 80 \%$ if possible. Pre-processing, development and evaluation of the algorithm was performed with Python version 3.7 and PyTorch 1.6 on Nvidia Tesla V100.

\subsubsection{Diagnostic Performances on Three Different External Test Datasets.} The diagnostic performances of our model on three different external test datasets are provided in Table \ref{table:various_ext}. Our model shows average AUCs of 0.941, 0.909, 0.915, average sensitivities of 87.0\%, 85.1\%, 85.9\%, average speicificities of 91.2\%. 84.7\%, 84.8\%, and average accuracy of 86.4\%, 85.9\%, 85.2\% in the three different institutional datasets obtained with different decvices and settings. These results demonstrate that our model retains stable performance (AUC $\geq 0.900$) irrespective of external settings, which confirms fair excellent generalization ability of our model.

\subsubsection{Comparison with baseline and SOTA models.} To compare the diagnostic performance of our model with the baseline and SOTA models, ResNet-50 was used as baseline and ViT (ViT-B/16), hybrid ViT model (R50-ViT-B/16) were used as SOTA models. All models except ours were trained using the pre-trained ImageNet weights since it significantly improves their performance, and were subjected to same training setting with our model for fair comparison. As shown in Table \ref{table:various_ext}, our model outperformed the SOTA models as well as the baseline in all of the external test datasets, demonstrating the superior performance and stability in real-world application.

\begin{table}[!t]
  \caption{Diagnostic performance the proposed model in various external test datasets from three different institutions and Comparison wiht baseline and SOTA methods.}
  \label{table:various_ext}
  \centering
  \resizebox{\textwidth}{!}
  {%
  \begin{tabular}{M{1.17cm}|M{1.12cm}|M{1.1cm}|M{1.15cm}|M{1.1cm}|M{1.12cm}|M{1.1cm}|M{1.15cm}|M{1.1cm}|M{1.12cm}|M{1.1cm}|M{1.15cm}|M{1.1cm}}
  \hline
  \multicolumn{1}{l|}{\textbf{Metrics}} & \multicolumn{4}{c|}{\textbf{External set 1 (CNUH)}} & \multicolumn{4}{c|}{\textbf{External set 2 (YNU)}} & \multicolumn{4}{c}{\textbf{External set 3 (KNUH)}} \\
  \cline{2-13}
  & {Avg.} & {Normal} & {Others} & {COVID} & {Avg.} & {Normal} & {Others} & {COVID} & {Avg.} & {Normal} & {Others} & {COVID} \\
  \hline
   \multicolumn{1}{l|}{AUC} & 0.941 & 0.927 & 0.948 & 0.947 & 0.909 & 0.938 & 0.955 & 0.833 & 0.915 & 0.938 & 0.938 & 0.868 \\
  \multicolumn{1}{l|}{Sensitivity} & 87.0 & 88.4 & 87.2 & 85.5 & 85.1 & 90.7 & 89.6 & 75.0 & 85.9 & 90.3 & 89.0 & 78.5 \\
   \multicolumn{1}{l|}{Specificity} & 91.2 & 88.9 & 84.7 & 100.0 & 84.7 & 86.2 & 87.7 & 80.2 & 84.8 & 85.3 & 89.2 & 80.0 \\
   \multicolumn{1}{l|}{Accuracy} & 86.4 & 88.5 & 84.9 & 85.8 & 85.9 & 89.2 & 88.3 & 80.1 & 85.2 & 87.8 & 89.1 & 78.7 \\
   \hline
   \hline
   \multicolumn{1}{l|}{\textbf{Methods}} & \multicolumn{4}{c|}{\textbf{AUC for External set 1}} & \multicolumn{4}{c|}{\textbf{AUC for External set 2}} & \multicolumn{4}{c}{\textbf{AUC for External set 3}} 
  \\
  \cline{2-13}
  & Avg. & {Normal} & {Others} & {COVID} & Avg. & {Normal} & {Others} & {COVID} &Avg. & {Normal} & {Others} & {COVID} \\
  \hline
   \multicolumn{1}{l|}{ResNet-50} & 0.856 & 0.859 & 0.910 & 0.799 & 0.856 & 0.826 & 0.943 & 0.800 & 0.876 & 0.869 & 0.849 & \textbf{0.910} \\
  \multicolumn{1}{l|}{ViT} & 0.857 & 0.856 & 0.865 & 0.851 & 0.852 & 0.833 & 0.887 & \textbf{0.837} & 0.811 & 0.843 & 0.686 & 0.904 \\
   \multicolumn{1}{l|}{ViT(hybrid)} & 0.890 & 0.893 & 0.903 & 0.875 & 0.890 & 0.913 & 0.942 & 0.815 & 0.863 & 0.927 & 0.799 & 0.864 \\
   \multicolumn{1}{l|}{\textbf{Ours}} & \textbf{0.941} & \textbf{0.927} & \textbf{0.948} & \textbf{0.947} & \textbf{0.909} & \textbf{0.938} & \textbf{0.955} & 0.833 & \textbf{0.915} & \textbf{0.938} & \textbf{0.938} & 0.868 \\
   \cline{1-13}
   \hline
  \end{tabular}}  
\end{table}



\subsubsection{Model Interpretability Results.} 
\begin{figure}[!b]
\includegraphics[width=\textwidth]{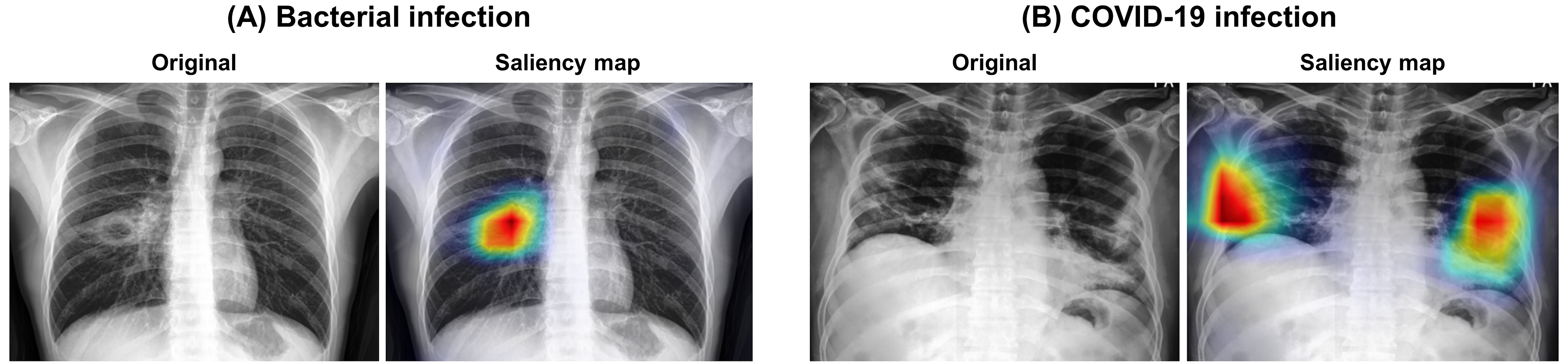}
\caption{Examples of visualization of saliency map for each disease classes. (A) Bacterial infection and (B) COVID-19 infection.} \label{fig2}
\end{figure}

Fig. \ref{fig2}  illustrates the examples of visualization of saliency map for each disease classes. As shown in Fig. \ref{fig2} (A), the our model well-localized a focal cavitary lesion caused by bacterial infection, while it was also able to delineate the multi-focal areas of involvement by virus which is common for COVID-19 infection as in Fig. \ref{fig2} (B).

\subsubsection{Self-supervised Contrastive Pre-training.} Since it has been suggested that the Transformer-based model may benefit from pre-training to learn the sequence structure by self-supervised manner before the fine-tuning for downstream tasks \cite{devlin2018bert, radford2018improving, chen2020generative}, we evaluated the benefit of self-supervised pre-training in Transformer-based SOTA (hybrid ViT) model and our model. We used SimCLR \cite{chen2020simple}, which is a self-supervised contrastive learning method with data augmentation framework, as our self-supervised pre-training method. As provided in Table \ref{table:comparison_ablation}, the experimental results reveal that the self-supervised pre-trainig was superfluous or even detrimental to our model since it already equipped with well-trained backbone network, though it slightly improved the performance of Transformer-based SOTA model. However, our model still outperformed SOTA model, which alludes that the corpus generated by the ResNet may not be a optimal input feature embedding for CXR classification.

\subsubsection{Ablation.} We conducted an ablation study to determine whether to have backbone network fixed or trainable after training with CheXpert dataset. The results of Table \ref{table:comparison_ablation} suggests that having weights of trainable is better than fixing them for in all of the 3 external test datasets, which is thought to be mitigated though improved capacity by trainable parameter of backbone network, dispelling the concern about overfitting.

\begin{table}[!t]
  \caption{Performance with or without self-supervised contrastive pre-training and Ablation study with fixed and trainable backbone weights.}
  \label{table:comparison_ablation}
  \centering
  \resizebox{\textwidth}{!}
  {%
  \begin{tabular}{M{1cm}|M{2cm}|M{2}|M{2cm}}
  \hline
  \multicolumn{1}{l|}{\textbf{Methods}} & \multicolumn{1}{c|}{\textbf{External 1 (CNUH)}} & \multicolumn{1}{c|}{\textbf{External 2 (YNU)}} & \multicolumn{1}{c}{\textbf{External 3 (KNUH)}} \\
  \cline{2-4}
  & \multicolumn{1}{c|}{Avg. AUC of 3 classes} & \multicolumn{1}{c|}{Avg. AUC of 3 classes} & \multicolumn{1}{c}{Avg. AUC of 3 classes} \\
  \hline
   \multicolumn{1}{l|}{ViT(hybrid) w/o pretrain} & \multicolumn{1}{c|}{0.890} & \multicolumn{1}{c|}{0.890}  & \multicolumn{1}{c}{0.863} \\
   \multicolumn{1}{l|}{ViT(hybrid) w pretrain} & \multicolumn{1}{c|}{0.881} & \multicolumn{1}{c|}{\textbf{0.911}}  & \multicolumn{1}{c}{0.870} \\
   \multicolumn{1}{l|}{\textbf{Ours w/o pretrain}} & \multicolumn{1}{c|}{\textbf{0.941}} & \multicolumn{1}{c|}{0.909}  & \multicolumn{1}{c}{\textbf{0.915}} \\
   \multicolumn{1}{l|}{Ours w pretrain} & \multicolumn{1}{c|}{0.935} & \multicolumn{1}{c|}{0.876}  & \multicolumn{1}{c}{0.903}  \\
   \hline
   \multicolumn{1}{l|}{Not trainable backbone} & \multicolumn{1}{c|}{0.908} & \multicolumn{1}{c|}{0.895}  & \multicolumn{1}{c}{0.899} \\
   \multicolumn{1}{l|}{\textbf{Trainable backbone}} & \multicolumn{1}{c|}{\textbf{0.941}} & \multicolumn{1}{c|}{\textbf{0.909}}  & \multicolumn{1}{c}{\textbf{0.915}}  \\
   \cline{1-4}
   \hline
  \end{tabular}}  
\end{table}

\section{Conclusion} In this work, we proposed a novel Vision Transformer model for COVID-19 CXR diagnosis, using the low-level CXR feature corpus. The novelty of this study lies in leveraging a backbone network trained to find low-level abnormal CXR findings in prebuilt large-scale dataset to embed feature corpus suitable for high-level disease classification with Transformer model. The experimental results on various external test datasets confirm that our model not only achieves SOTA performance in the diagnosis of COVID-19 and other infectious disease but also retains stable performance irrespecitve of the external settings, which is a sine-qua-non for widespread application of system. In addition, we provided interpretable results with improved visualization method beyond attention, which is expected to be of great help to clinicians.

\bibliographystyle{splncs04}
\bibliography{COVID-19_bib}

\begin{thebibliography}{10}
\providecommand{\url}[1]{\texttt{#1}}
\providecommand{\urlprefix}{URL }
\providecommand{\doi}[1]{https://doi.org/#1}

\bibitem{apostolopoulos2020covid}
Apostolopoulos, I.D., Mpesiana, T.A.: Covid-19: automatic detection from x-ray
  images utilizing transfer learning with convolutional neural networks.
  Physical and Engineering Sciences in Medicine  \textbf{43}(2),  635--640
  (2020)

\bibitem{bernheim2020chest}
Bernheim, A., Mei, X., Huang, M., Yang, Y., Fayad, Z.A., Zhang, N., Diao, K.,
  Lin, B., Zhu, X., Li, K., et~al.: Chest ct findings in coronavirus disease-19
  (covid-19): relationship to duration of infection. Radiology p. 200463 (2020)

\bibitem{chefer2020transformer}
Chefer, H., Gur, S., Wolf, L.: Transformer interpretability beyond attention
  visualization. arXiv preprint arXiv:2012.09838  (2020)

\bibitem{chen2020more}
Chen, L., Min, Y., Zhang, M., Karbasi, A.: More data can expand the
  generalization gap between adversarially robust and standard models. In:
  International Conference on Machine Learning. pp. 1670--1680. PMLR (2020)

\bibitem{chen2020generative}
Chen, M., Radford, A., Child, R., Wu, J., Jun, H., Luan, D., Sutskever, I.:
  Generative pretraining from pixels. In: International Conference on Machine
  Learning. pp. 1691--1703. PMLR (2020)

\bibitem{chen2020simple}
Chen, T., Kornblith, S., Norouzi, M., Hinton, G.: A simple framework for
  contrastive learning of visual representations. In: International conference
  on machine learning. pp. 1597--1607. PMLR (2020)

\bibitem{cozzi2020chest}
Cozzi, D., Albanesi, M., Cavigli, E., Moroni, C., Bindi, A., Luvar{\`a}, S.,
  Lucarini, S., Busoni, S., Mazzoni, L.N., Miele, V.: Chest x-ray in new
  coronavirus disease 2019 (covid-19) infection: findings and correlation with
  clinical outcome. La radiologia medica  \textbf{125},  730--737 (2020)

\bibitem{de2020bimcv}
De~La Iglesia~Vay{\'a}, M., Saborit, J.M., Montell, J.A., Pertusa, A., Bustos,
  A., Cazorla, M., Galant, J., Barber, X., Orozco-Beltr{\'a}n, D.,
  Garc{\'\i}a-Garc{\'\i}a, F., et~al.: Bimcv covid-19+: a large annotated
  dataset of rx and ct images from covid-19 patients. arXiv preprint
  arXiv:2006.01174  (2020)

\bibitem{devlin2018bert}
Devlin, J., Chang, M.W., Lee, K., Toutanova, K.: Bert: Pre-training of deep
  bidirectional transformers for language understanding. arXiv preprint
  arXiv:1810.04805  (2018)

\bibitem{dosovitskiy2020image}
Dosovitskiy, A., Beyer, L., Kolesnikov, A., Weissenborn, D., Zhai, X.,
  Unterthiner, T., Dehghani, M., Minderer, M., Heigold, G., Gelly, S., et~al.:
  An image is worth 16x16 words: Transformers for image recognition at scale.
  arXiv preprint arXiv:2010.11929  (2020)

\bibitem{hemdan2020covidx}
Hemdan, E.E.D., Shouman, M.A., Karar, M.E.: Covidx-net: A framework of deep
  learning classifiers to diagnose covid-19 in x-ray images. arXiv preprint
  arXiv:2003.11055  (2020)

\bibitem{hu2020challenges}
Hu, Y., Jacob, J., Parker, G.J., Hawkes, D.J., Hurst, J.R., Stoyanov, D.: The
  challenges of deploying artificial intelligence models in a rapidly evolving
  pandemic. Nature Machine Intelligence  \textbf{2}(6),  298--300 (2020)

\bibitem{irvin2019chexpert}
Irvin, J., Rajpurkar, P., Ko, M., Yu, Y., Ciurea-Ilcus, S., Chute, C.,
  Marklund, H., Haghgoo, B., Ball, R., Shpanskaya, K., et~al.: Chexpert: A
  large chest radiograph dataset with uncertainty labels and expert comparison.
  In: Proceedings of the AAAI Conference on Artificial Intelligence. vol.~33,
  pp. 590--597 (2019)

\bibitem{narin2020automatic}
Narin, A., Kaya, C., Pamuk, Z.: Automatic detection of coronavirus disease
  (covid-19) using x-ray images and deep convolutional neural networks. arXiv
  preprint arXiv:2003.10849  (2020)

\bibitem{ng2020covid}
Ng, K., Poon, B.H., Kiat~Puar, T.H., Shan~Quah, J.L., Loh, W.J., Wong, Y.J.,
  Tan, T.Y., Raghuram, J.: Covid-19 and the risk to health care workers: a case
  report. Annals of internal medicine  \textbf{172}(11),  766--767 (2020)

\bibitem{oh2020deep}
Oh, Y., Park, S., Ye, J.C.: Deep learning covid-19 features on cxr using
  limited training data sets. IEEE Transactions on Medical Imaging
  \textbf{39}(8),  2688--2700 (2020)

\bibitem{radford2018improving}
Radford, A., Narasimhan, K., Salimans, T., Sutskever, I.: Improving language
  understanding by generative pre-training  (2018)

\bibitem{shi2020radiological}
Shi, H., Han, X., Jiang, N., Cao, Y., Alwalid, O., Gu, J., Fan, Y., Zheng, C.:
  Radiological findings from 81 patients with covid-19 pneumonia in wuhan,
  china: a descriptive study. The Lancet infectious diseases  \textbf{20}(4),
  425--434 (2020)

\bibitem{signoroni2020end}
Signoroni, A., Savardi, M., Benini, S., Adami, N., Leonardi, R., Gibellini, P.,
  Vaccher, F., Ravanelli, M., Borghesi, A., Maroldi, R., et~al.: End-to-end
  learning for semiquantitative rating of covid-19 severity on chest x-rays.
  arXiv preprint arXiv:2006.04603  (2020)

\bibitem{tahamtan2020real}
Tahamtan, A., Ardebili, A.: Real-time rt-pcr in covid-19 detection: issues
  affecting the results. Expert review of molecular diagnostics
  \textbf{20}(5),  453--454 (2020)

\bibitem{vaswani2017attention}
Vaswani, A., Shazeer, N., Parmar, N., Uszkoreit, J., Jones, L., Gomez, A.N.,
  Kaiser, L., Polosukhin, I.: Attention is all you need. arXiv preprint
  arXiv:1706.03762  (2017)

\bibitem{wang2020covid}
Wang, L., Lin, Z.Q., Wong, A.: Covid-net: A tailored deep convolutional neural
  network design for detection of covid-19 cases from chest x-ray images.
  Scientific Reports  \textbf{10}(1),  1--12 (2020)

\bibitem{wang2017chestx}
Wang, X., Peng, Y., Lu, L., Lu, Z., Bagheri, M., Summers, R.M.: Chestx-ray8:
  Hospital-scale chest x-ray database and benchmarks on weakly-supervised
  classification and localization of common thorax diseases. In: Proceedings of
  the IEEE conference on computer vision and pattern recognition. pp.
  2097--2106 (2017)

\bibitem{wang2020weakly}
Wang, X., Deng, X., Fu, Q., Zhou, Q., Feng, J., Ma, H., Liu, W., Zheng, C.: A
  weakly-supervised framework for covid-19 classification and lesion
  localization from chest ct. IEEE transactions on medical imaging
  \textbf{39}(8),  2615--2625 (2020)

\bibitem{wong2020frequency}
Wong, H.Y.F., Lam, H.Y.S., Fong, A.H.T., Leung, S.T., Chin, T.W.Y., Lo, C.S.Y.,
  Lui, M.M.S., Lee, J.C.Y., Chiu, K.W.H., Chung, T.W.H., et~al.: Frequency and
  distribution of chest radiographic findings in patients positive for
  covid-19. Radiology  \textbf{296}(2),  E72--E78 (2020)

\bibitem{ye2020weakly}
Ye, W., Yao, J., Xue, H., Li, Y.: Weakly supervised lesion localization with
  probabilistic-cam pooling. arXiv preprint arXiv:2005.14480  (2020)

\bibitem{zech2018variable}
Zech, J.R., Badgeley, M.A., Liu, M., Costa, A.B., Titano, J.J., Oermann, E.K.:
  Variable generalization performance of a deep learning model to detect
  pneumonia in chest radiographs: a cross-sectional study. PLoS medicine
  \textbf{15}(11),  e1002683 (2018)

\bibitem{zhang2020covid}
Zhang, J., Xie, Y., Li, Y., Shen, C., Xia, Y.: Covid-19 screening on chest
  x-ray images using deep learning based anomaly detection. arXiv preprint
  arXiv:2003.12338  (2020)

\bibitem{zheng2020deep}
Zheng, C., Deng, X., Fu, Q., Zhou, Q., Feng, J., Ma, H., Liu, W., Wang, X.:
  Deep learning-based detection for covid-19 from chest ct using weak label.
  MedRxiv  (2020)

\end{thebibliography}

\end{document}